\documentstyle[mprocl,epsfig]{article}

\input{rotate}

\def\Journal#1#2#3#4{{#1} {\bf #2}, #3 (#4)}

\def\NPA{{\em Nucl. Phys.} A}
\def\NPB{{\em Nucl. Phys.} B}
\def\PLB{{\em Phys. Lett.}  B}
\def\PRL{\em Phys. Rev. Lett.}
\def\PRD{{\em Phys. Rev.} D}
\def\PRC{{\em Phys. Rev.} C}

\def\be{\begin{equation}}
\def\ee{\end{equation}}
\def\bea{\begin{eqnarray}}
\def\eea{\end{eqnarray}}

\newcommand{\xbf}[1]{\mbox{\boldmath $ #1 $}}

\newcommand{\Gen}{\mbox{G$_{C}^{n}$}}

\newcommand{\GCndelta}{\mbox{G$_{\mathrm{C}2}^{N \to \Delta}$}}

\newcommand{\etal}{\mbox{\it et~al.}}

\begin{document}

\title{Relations between $N$ and $\Delta$ electromagnetic form factors
\footnote{Excited Nucleons and Hadronic Structure, Proceedings of the 
NSTAR 2000 conference, 
Eds. V. D. Burkert, L.Elouadrhiri, J.J. Kelly,
R. C. Minehart, World Scientific, Singapore, 2001, pg. 59}
}

\author{A. J. Buchmann}

\address{Institute for Theoretical Physics, University of T\"ubingen, 
D-72076 T\"ubingen,\\ Germany\\E-mail: alfons.buchmann@uni-tuebingen.de}   


\maketitle\abstracts{ 
The inclusion of two-body exchange currents in the constituent quark model
leads to several new relations between the electromagnetic form factors of 
nucleon and $\Delta(1232)$. These are: (i) the neutron charge form factor
can be expressed as the difference between proton and $\Delta^+$ 
charge form factors, and (ii) the $N \to \Delta$ charge quadrupole $(C2)$
transition form factor is connected to the
charge monopole $(C0)$ form factor of the neutron. The latter
relation is used to estimate the charge radius of constituent quarks. 
Furthermore, we find that exchange currents
do {\it not} modify the $SU(6)$ relation between the magnetic 
$N \to \Delta$ and the magnetic neutron
form factor. Consequently, after including exchange currents,
the $C2/M1$ ratio in the $N \to \Delta$ transition can be expressed as a 
ratio of the elastic charge and magnetic neutron form factors as 
follows 
$$
\frac{C2}{M1}({\bf q}^2) = \frac{M_N}{2 \sqrt{{\bf q}^2}} 
\frac{G_{C}^{n}({\bf q}^2)}{G_M^{n}({\bf q}^2)} \, .  
$$
}

\section{Introduction}

Baryons are complicated many-particle systems composed of valence
quarks, which carry the quantum numbers, and 
nonvalence quark degrees of freedom, such as 
quark-antiquark ($q \bar q$) pairs and gluons. 
The constituent quark model with two-body exchange currents
describes both these aspects baryon structure~\cite{Buc91}.  One-body
currents describe the interaction of the photon with one valence quark at
a time.  Two-body exchange currents are connected with the exchange
particles and with $q \bar q$ pairs.
Baryon properties which are dominated by
two-body exchange currents show their common dynamical origin in
analytical interrelations between them.

A new quark model relation between the neutron 
charge form factor \Gen\ and the quadrupole transition form factor \GCndelta\ 
is used in order to predict \GCndelta\
and the C2/M1 ratio from the elastic neutron form factor data.
An astonishingly good agreement with the 
direct pion electroproduction data is found~\cite{Gra00}.

\section{$N$ and $\Delta$ charge monopole form factors} 

\begin{figure}[h]
$$\mbox{
\epsfxsize 11.5 true cm
\epsfysize 4.0 true cm
\setbox0= \vbox{
\hbox { \centerline{
\epsfbox{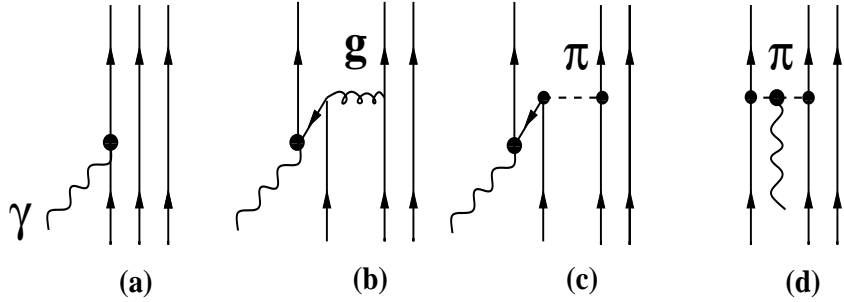}
}} 
} 
\box0
} $$
\vspace{-0.8cm}
\caption[Exchange currents]{Feynman diagrams of the four vector
current $J^{\mu}=(\rho, {\bf J})$:
photon ($\gamma$) coupling to (a) one-body current $J^{\mu}_{[1]}$, 
and to (b-d) two-body gluon and pion exchange currents $J^{\mu}_{[2]}$.
Diagrams (b-d) must be taken into account in order to satisfy the continuity
equation $q_{\mu}J^{\mu}=0$ for the electromagnetic current 
$J_{\mu}$. They represent the nonvalence (gluon and pion) degrees of 
freedom in the nucleon in the presence of an external electromagnetic field.}
\label{fig:feynmec}
\end{figure}

In a quark potential model with gluon and pion 
exchange currents the baryon charge consists of a sum of 
one- and two-quark pieces: $\rho({\bf q})= \rho_{[1]}({\bf q}) + 
\rho_{[2]}({\bf q})$. After a multipole expansion up to quadrupole terms, 
the one- and two-body quark operators corresponding to
Fig.1 can be schematically written as
\bea
\label{eq:structure}
\rho_{[1]}({\bf  q}) & \approx & 
[Y^0({\bf r}_i) \times Y^0({\bf q})]^0
-\sqrt{5}  [Y^2({\bf r}_i) \times Y^2({\bf q})]^0 \nonumber \\
\rho_{[2]}({\bf q}) & \approx & 
[[\xbf{\sigma}_i \times \xbf{\sigma}_j]^0 \times Y^0({\bf q})]^0
+\frac{1}{\sqrt{2}}[\, [\xbf{\sigma}_i \times \xbf{\sigma}_j]^2 \times
Y^2({\bf q})\, ]^0.
\eea 
where ${\bf r}_i$ is the spatial, and $\xbf{\sigma}_i$ the spin operator 
of a single quark, 
${\bf q}$ is the three-momentum transfer of the photon, and
$Y^l$ a spherical harmonic of rank $l$.
The spin-dependent two-body terms come from the exchange current diagrams in
Fig.~\ref{fig:feynmec}(b-d). 

Evaluating these operators between three-quark proton wave functions, one 
obtains for the proton charge radius \cite{Buc91}
\be
\label{rpmec}
r^2_{p}   =  b^2 + r_{\gamma q}^2+ 
{b^2\over 2 m_q}  
\bigl ( \delta_g(b) - \delta_{\pi}(b) \bigr ).
\ee
Here, $b$ is the quark core (matter) radius of the nucleon, 
$r_{\gamma q}$ the finite charge radius of the constituent quark,
and $m_q$ the constituent quark mass satisfying $M_N=3 m_q$.
The terms proportional to $\delta_g(b)$ and $\delta_{\pi}(b)$ 
describe the gluon and pion exchange current contributions to 
the charge radius. These functions also express 
the gluon and pion contributions to the $N$-$\Delta$ 
mass splitting:
$
M_{\Delta}-M_N=\delta_g(b)+\delta_{\pi}(b).
$ 
The proton charge radius 
is mainly determined by the  
valence quark terms $b^2$ and $r_{\gamma q}^2$, i.e., 
the one-body current depicted in Fig.\ref{fig:feynmec}(a);
the exchange currents of Fig.\ref{fig:feynmec}(b-c) provide
only a small correction to $r^2_{p}$.

\begin{figure}[t]
$$\hspace{0.2cm} \mbox{
\epsfxsize 9.0 true cm
\epsfysize 11.5 true cm
\setbox0= \vbox{
\hbox {
\epsfbox{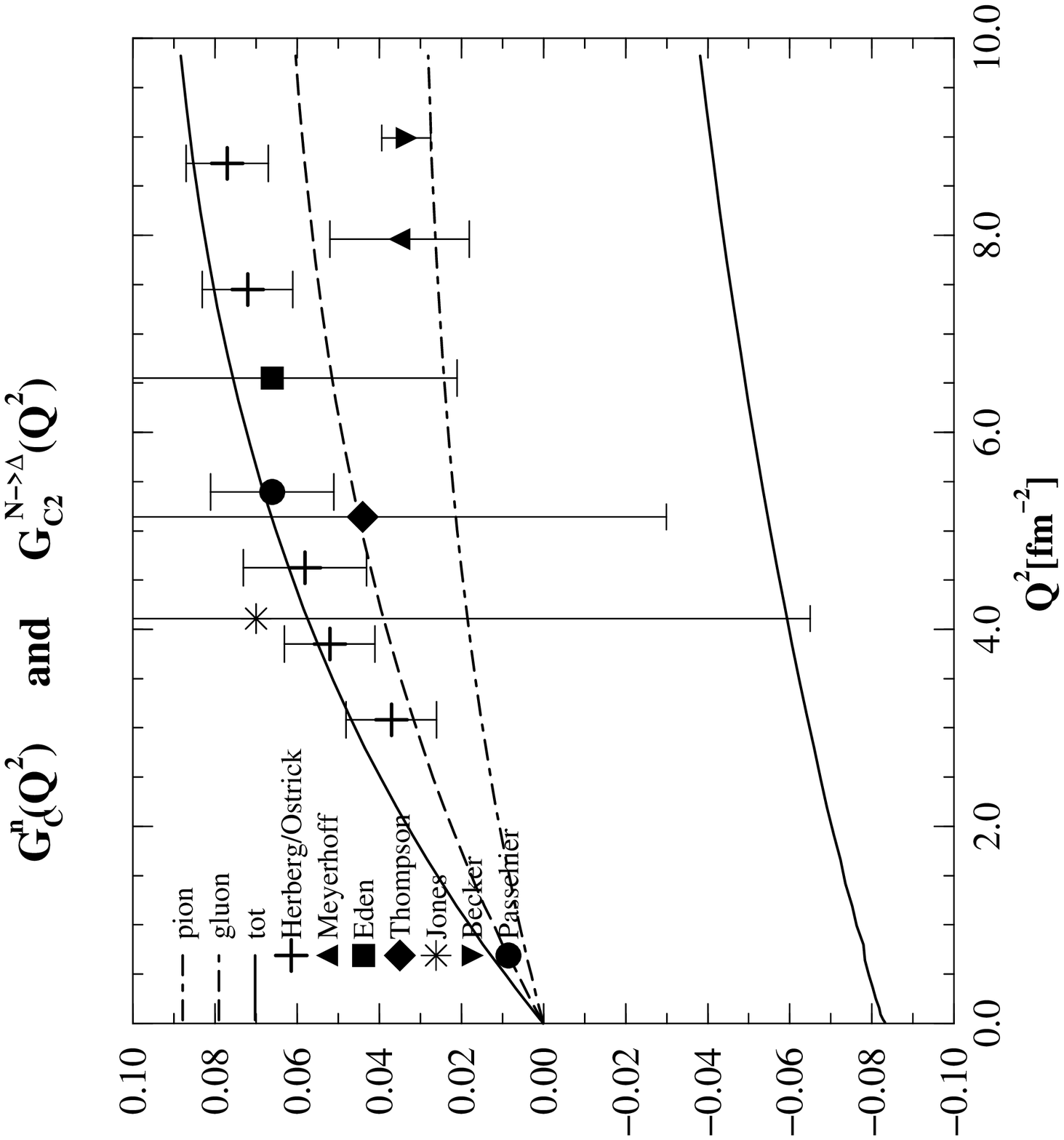}
} 
} 
\rotr0
} $$
\vspace{-0.8cm}
\caption[Interpretation]{
Neutron charge form factor $G_C^n(Q^2)$ (upper curves) and the
$N \to \Delta$ quadrupole transition form factor 
$G_{C2}^{p \to \Delta^+}(Q^2)$ (lower curve) 
as a function of four-momentum transfer $Q^2=-q^2$ as predicted 
quark model with exchange currents \cite{Buc91,Buc97b}. 
The lower curve is 
$G_{C2}^{p \to \Delta^+}(Q^2)=-3\sqrt{2}\, \, G_C^n(Q^2)/Q^2$\, as
predicted by the quark model with exchange currents~\cite{Buc00a}.
The data are from Ref.~\cite{Her99}.}
\label{nchaff99}
\end{figure}

In contrast, the Sachs charge form factor of the neutron $G_C^n(q^2)$ 
(see Fig.\ref{nchaff99}) 
and the corresponding charge radius $r_n^2=-6 (d/d{\bf q}^2) 
G_C^n({\bf q}^2) \mid_{{\bf q}^2=0}$  are dominated by the 
quark-antiquark pair exchange currents, and one obtains~\cite{Buc91}
\be
\label{eq:rnmec}
r^2_{n}   =  -{b^2\over 3m_q} \bigl ( \delta_g(b)+ \delta_{\pi}(b) \bigr )
=-b^2  \left ( \frac{M_{\Delta}-M_N}{M_N} \right ). 
\ee
The above relations show that the spin-dependent gluon and pion exchange
potentials, which generate the $N$-$\Delta$ mass splitting, are
also responsible for the nonvanishing neutron charge radius via the
corresponding spin-spin term in Eq.(\ref{eq:structure}).
The valence quark terms $b^2$ and $r_{\gamma q}^2$ 
do not appear in Eq.(\ref{eq:rnmec}). The reasons for this will become 
clear soon. 

The charge radii of all charged $\Delta$(1232) states are calculated 
in the same $N_c=3$ quark model as~\cite{Buc97}  
\be
\label{rdel}
r^2_{\Delta} = b^2 + r^2_{\gamma q}+ {b^2\over 6 m_q} (5\delta_g(b) -
\delta_{\pi}(b)), \quad r^2_{\Delta^0}=0.
\ee
The charge form factor of the $\Delta^0$ and the corresponding 
charge radius are zero (for $N_c=3$) 
as it should be on general grounds \cite{Col85}. Subtracting 
Eq.(\ref{rdel}) from Eq.(\ref{rpmec})  and Eq.(\ref{eq:rnmec})   
the valence quark contributions $b^2$ and $r_{\gamma q}^2$ cancel,  
and one finds~\cite{Buc97} 
\be
\label{crdn}
r^2_p-r^2_{\Delta^+} = r^2_n \, , \qquad r^2_n-r^2_{\Delta^0} = r^2_n . 
\ee

Eqs.(\ref{crdn}) are the first moments
of the more general 
relations~\cite{Buc00a} between the charge form factors of the 
$N$ and $\Delta$
\be
\label{crdnff}
G_C^{p}({\bf q}^2)-G_C^{\Delta^+}({\bf q}^2)= G_C^{n}({\bf q}^2) \, ,
\qquad
G_C^{n}({\bf q}^2)-G_C^{\Delta^0}({\bf q}^2)= G_C^{n}({\bf q}^2) \, .
\ee
Including the $\Delta^{++}$ and the $\Delta^-$ charge states 
this can be written in closed form as
\be
\label{crdnff2}
G_C^{\Delta}({\bf q}^2)=\left ( G_C^{p}({\bf q}^2)- G_C^{n}({\bf q}^2)\right )
e_{\Delta},
\ee 
where $e_{\Delta}=(1+2 T_3)/2$ is the $\Delta$ charge.
These relations are {\it not}  equivalent to the  
$SU(6)$ result 
$G_C^{\Delta}=(G_C^p+G_C^n)/2 +(G_C^p-G_C^n) T_3$, where $T_3$ is the 
third component of the $\Delta$ isospin.          
Eq.(\ref{crdnff2}) contains the important symmetry breaking effect
coming from the spin-spin term in Eq.(\ref{eq:structure}).

Dillon and Morpurgo \cite{Mor99} have 
recently shown that Eq.(\ref{crdn}) is a direct
consequence of the underlying $SU(6)$ spin-flavor symmetry 
{\it and } the quark-gluon dynamics of quantum chromodynamics. 
They have also shown that 
three-body currents slightly modify, but do not invalidate the general
relationship between the proton, neutron, and $\Delta$ charge radii.
The work of Dillon and Morpurgo makes it clear that Eq.(\ref{crdn}) 
originally found in the quark model with exchange currents, 
is a general relation if three-body operators and
strange quark loops are neglected.

Considering constituent quarks with an arbitrary 
number of colors $N_c$, one can generalize these 
findings~\cite{Buc00b}. Eq.(\ref{crdn}) then no longer
holds for arbitrary $N_c$, but the relation
\be
\label{crdn1}
r^2_p-r^2_{\Delta^+} =r^2_n-r^2_{\Delta^0}
\ee
is valid for {\it any} $N_c$.
It is broken only by three-body $O(1/N_c^2)$ terms,
which are suppressed compared to the two-body terms included.
For $N_c=3$, $r_{\Delta^0}^2=0$ and we reobtain Eq.(\ref{crdn}). 
The general $N_c$ analysis gives the $N-\Delta$ charge radius
relationship a rigorous theoretical foundation~\cite{Buc00b}. 

The following discussion suggests a 
connection between the $N-\Delta$ mass and charge radius difference.
We recall that the spin-spin structure 
$\xbf{\sigma}_i \cdot \xbf{\sigma}_j$ in 
Eq.(\ref{eq:structure})
is responsible for the splitting between $N$ and  $\Delta$ charge radii.
It leads to a $\Delta$ charge radius that is 
larger than the proton charge radius by an amount that is equal to the 
negative neutron charge radius. 
The neutron charge radius is nonzero, because the spin-spin term
in Eq.(\ref{eq:structure}) gives different matrix elements for 
quark pairs in spin~0 and spin~1 states.
This splitting 
of the $N$ and $\Delta$ charge radii is of the same generality as,
and closely connected with the $N-\Delta$ mass splitting due to the 
spin-spin interaction in the Hamiltonian. The latter is repulsive 
in quark pairs with spin 1 and makes the $\Delta$ heavier than the nucleon. 
Combining Eq.(\ref{eq:rnmec}) and Eq.(\ref{crdn}) and the fact
that baryon charge radii and masses have the same large $N_c$ operator
expansion \cite{Buc00b} we conjecture that 
\be
\label{charge mass}
\frac{r^2_{\Delta^+}-r_p^2}{r^2_{\Delta^0}-r_n^2}
 =\frac{M_{\Delta^+}-M_p}{M_{\Delta^0}-M_n}
\ee
contains some of the three-body corrections not included in Eq.(\ref{crdn1}).
%
\goodbreak
\begin{table}[htb]
\caption[Magnetic moments]{$\Delta$(1232) magnetic moments
based on the relations suggested in this paper.
As input the experimental proton and neutron
magnetic moments are used. The  experimental range for the
$\Delta^{++}$ magnetic moment is $\mu_{\Delta^{++}}=(3.7-7.5)$ $\mu_N$ 
\cite{PDG94}. 
For the $N \to \Delta$ transition magnetic moment experimental values
lie between $\mu_{p \to \Delta^+}=(3.5-4.2)$ $\mu_N$ \cite{Buc97b}. 
All entries are in given in units of nuclear magnetons $\mu_N=e/(2M_p)$.  }
\begin{center}
\nobreak
\begin{tabular}[h]{| l | r | r |} 
\hline
Baryon  & Quark Model & $SU(6)$  \\  \hline
$ \Delta^{++} $ & $ 5.28 $ & $5.58$ \\
$ \Delta^{+} $ & $ 2.64 $ & $2.79$ \\ 
$ \Delta^{0} $ & $ 0.000 $ & $0.000 $  \\ 
$ \Delta^{-} $ & $ -2.64 $ & $-2.79$  \\
$ p\to \Delta^{+} $ & $ 2.70 $ & $ 2.70 $  \\
$ n\to \Delta^{0} $ & $ 2.70 $ & $ 2.70 $  \\
\hline
\end{tabular} 
\end{center}
\end{table}

\goodbreak 
\section{$N$ and $\Delta$ magnetic dipole form factors} 
\nobreak

The quark model with two-body exchange currents also
relates the magnetic form factors of the $N$ and  $\Delta$
\be
\label{relmag}
G_M^{\Delta}({\bf q}^2) = 3 
\left ( G_M^p({\bf q}^2) + G_M^n({\bf q}^2) \right ) e_{\Delta}.
\ee
Eq.(\ref{relmag}) differs from the $SU(6)$ relation~\cite{Beg64}:  
$G_M^{\Delta}=G_M^p e_{\Delta}$. 
There is no difference between the quark model with 
exchange currents and the $SU(6)$ result if the 
additional $SU(6)$
relation $G_M^n=-2G_M^p/3$ is used in Eq.(\ref{relmag}).
Our predictions for the $\Delta$ magnetic moments 
based on Eq.(\ref{relmag}) are given in Table 1.
We observe that the $\Delta^+$ magnetic moment is 
only slightly smaller than the proton magnetic moment.

\goodbreak 
\section{$N \to \Delta$ charge quadrupole transition form factor}
\label{subsec:ndquad}
\nobreak

In the constituent quark model with exchange currents a connection
between the neutron charge form factor $G_C^n({\bf q}^2)$ 
and the $N\to\Delta$ quadrupole transition form factor 
$G_{C2}^{p \to \Delta^+}({\bf q}^2)$ emerges \cite{Buc00a}:
\be
\label{qnff} 
G_{C2}^{p \to \Delta^+}({\bf q}^2)= -{3 \sqrt{2}\over {\bf q}^2} 
G_C^n({\bf q}^2)= -{3 \sqrt{2}\over {\bf q}^2} 
\left ( G_C^p({\bf q}^2) - G_C^{\Delta^+}({\bf q}^2) \right ), 
\ee
which is plotted in Fig.\ref{nchaff99} (lower curve).

In the low-momentum transfer limit we derive from Eq.(\ref{qnff})
that the $N \to \Delta$ transition quadrupole moment $Q_{p \to \Delta^+}$ 
is determined by the neutron charge radius~\cite{Buc97}
\be
\label{quad}
Q_{p \to \Delta} ={ r_n^2 \over \sqrt{2}}.
\ee
We recall that $Q_{p \to \Delta^+}$ is 
in combination with the $\Delta^+$ quadrupole moment $Q_{\Delta^+}$ 
a measure of the intrinsic deformation of the nucleon and $\Delta$.
The quantities that determine the intrinsic deformation
are the intrinsic quadrupole moments $Q_0^p$ and $Q_0^{\Delta}$.
The connection between the observable (spectroscopic) 
 $Q_{p\to\Delta^+}$ and the
$Q_{\Delta^+}$ and corresponding intrinsic quadrupole moments 
has recently been evaluated in different models~\cite{Buc00c}. 
In the quark model we find using the empirical neutron charge 
radius~\cite{kopecky} 
$Q_0^p = -Q_0^{\Delta^+}=-Q_{\Delta^+} = -\sqrt{2} Q_{p \to \Delta}= 
-r_n^2=+0.113$ fm$^2$. A negative $C2/M1$ ratio therefore implies  
a prolate (cigar-shaped)  intrinsic deformation of the nucleon and an
oblate (pancake-shaped) intrinsic deformation of the $\Delta$. 

The quark model with exchange currents explains $Q_{p \to \Delta^+}$
as a {\it double spin flip} of two quarks, with all valence quarks 
remaining in the dominant, spherically symmetric $L=0$ state. 
The spin-flip of {\it two} quarks comes from the tensor structure 
in Eq.(\ref{eq:structure}). 
The latter is closely related to the tensor term in the Hamiltonian,
which via the $D$ waves in the $N$ and $\Delta$ 
also contributes  to $Q_{p \to \Delta^+}$. This orbital excitation 
of a valence quark   
amounts to about 20$\%$ (due to the smallness of the $D$ wave amplitudes)
of the double spin flip amplitude~\cite{Buc97}.
We conclude that the {\it collective} $q{\bar q}$ degrees of freedom
are mainly responsible for the deformation of the $N$ and $\Delta$.
The importance of the spin tensor in Eq.(\ref{eq:structure}) 
for a complete explanation of the $N \to \Delta$ quadrupole transition
moment in the quark model was anticipated by Morpurgo \cite{Mor89}.

We have also calculated the radius of the $N \to \Delta$
transition quadrupole form factor~\cite{Buc97b},  
and obtained\footnote{The term 
${1 \over 4}b^2$ in Eq.(53) of Ref.\cite{Buc97b} should be replaced 
by ${11 \over 20} b^2$.}
\be
\label{quadrad}
r^{2}_{Q, \, p \to \Delta^+}= {11\over 20}\, b^2 +r^2_{\gamma q}.
\ee
Unlike in Eq.(\ref{rpmec}) 
there is no correction from two-body exchange currents in Eq.(\ref{quadrad}),
which makes the quadrupole transition radius an ideal observable 
to experimentally determine the quark charge radius $r_{\gamma q}$.
With the help of Eq.(\ref{qnff}) the quadrupole 
transition radius can be expressed 
as
\be
r^2_{Q,\, p \to \Delta^+}= ({18/r_n^2}) \, {(d/d{\bf q}^2)^2}
G_C^n({\bf q}^2)\bigl \vert_{{\bf q}^2=0}=\frac{3}{10}\frac{r_n^4}{r_n^2}, 
\ee 
where $r_n^4$ is the fourth moment of the neutron charge distribution.
Because the quark core radius $b$ is fixed by Eq.(\ref{eq:rnmec}), 
one can extract the charge radius of the light constituent 
quarks from the $G_C^n({\bf q}^2)$ data. 
A recent fit~\cite{Gra00} to the 
$G_C^n$  data determines the fourth moment of $G_C^n$ as
$r_n^4=-0.32$ fm$^4$ and the transition quadrupole radius as 
$r^2_{Q, \, p \to \Delta^+}=0.84(21)$fm$^2$.     
An additional data point of $G_C^n$ at $Q^2=0.9$ GeV$^{-2}$ would 
reduce the error by a factor of three. 
From Eq.(\ref{quadrad}) and Eq.(\ref{eq:rnmec}) we obtain 
$r^2_{\gamma q}=0.64$ fm$^2$, a rather large constituent quark charge radius.
This implies a proton charge radius of about 1 fm.

\goodbreak 
\section{$N \to \Delta$ magnetic dipole transition form factor}
\nobreak
After including
the gauge-invariant two-body exchange currents of Fig.\ref{fig:feynmec}(b-d), 
the $SU(6)$ Beg-Lee-Pais relation~\cite{Beg64}  between the 
magnetic $N\to \Delta$ transition and the neutron magnetic moments 
$\mu_{p \to \Delta^+}=-\sqrt{2} \mu_n$ remains unchanged, 
and holds even at finite momentum transfers
\be
\label{magff} 
G_M^{p \to \Delta^+}({\bf q}^2)= -\sqrt{2}\, G_M^n({\bf q}^2),  \qquad 
\mu_{p\to \Delta^+}= -\sqrt{2} \, \mu_n. 
\ee
The $N \to \Delta$ transition magnetic moment predicted by Eq.(\ref{magff}) 
underestimates the empirical value~\cite{Bec97}
$\mu_{p\to \Delta^+}^{exp}=(3.5-4.0) \, \mu_N$ by about $(30-50)\%$. 

On the other hand, replacing  $G_M^n$ with the help of 
Eq.(\ref{relmag}) allows to consider the problem from a different perspective
\be
\label{transmm}
G_M^{p \to \Delta^+}({\bf q}^2) 
= \sqrt{2} \left ( G_M^{p}({\bf q}^2) -
{ 1\over 3} \, G_M^{\Delta^+}({\bf q}^2) \right ), \quad 
\mu_{p\to \Delta^+}= \sqrt{2} 
\left ( \mu_p - \frac{1}{3} \mu_{\Delta^+} \right ).  
\ee
Eq.(\ref{transmm}), which describes the 
transition magnetic form factor in terms of the magnetic form factors of the
two baryons involved in the transition,  constraints  
$\mu_{p \to \Delta^+}$ to values  below $3 \mu_N$.
In order to make $\mu_{p\to \Delta^+}$ larger
one needs an unacceptably small or 
even a negative value for the $\Delta^+$ magnetic moment.
Thus, $\mu_{p \to \Delta^+}^{exp}$ most likely includes diagrams that
should not be included in the definition of          
the proper strength of the electromagnetic $\gamma N \Delta$ vertex. 
An analogous observation has been made for the $\pi N \Delta$ coupling 
strength~\cite{Buc99}. 

For the radius of the $N \to \Delta$ magnetic transition form factor,
$r^2_{M, \, p \to \Delta^+}$  we find
\be
r^2_{M, \, p \to \Delta^+}= \frac{-\sqrt{2}\, \mu_n}{\mu_{p \to \Delta^+}} 
\,\,  r^2_{M, \, n} \ , 
\ee
where 
$r^2_{M, \,  n}$ is the magnetic radius of the neutron.
With the $SU(6)$ result of Eq.(\ref{magff}), which is 
equivalent to the quark model result with two-body exchange currents, 
we obtain $r^2_{M, \, p \to \Delta^+}=r^2_{M, \, n}$.
If $\mu_{p \to \Delta^+}>-\sqrt{2}\,\mu_n$ we get 
$r^2_{M, \, p \to \Delta^+}<r^2_{M, \, n}$, contradicting the experimental
observation that the $N \to \Delta$ magnetic transition form factor drops 
off faster than the magnetic neutron form factor.

\goodbreak 
\section{$C2/M1$ ratio in the $N \to \Delta$ transition}
\nobreak

Combining Eq.({\ref{qnff}) and Eq.(\ref{magff}) we find that the 
ratio of the charge quadrupole and magnetic dipole 
$N\to \Delta$ transition form factors can be expressed 
in terms of the elastic neutron form factors~\cite{Buc00a}
\be
\label{c2m1}
\frac{C2}{M1}({\bf q^2}) = 
\frac{M_N}{2 \sqrt{{\bf q}^2} } {G_{C}^{n}({\bf q}^2)
\over G_M^{n}({\bf q}^2)}, 
\qquad
\frac{C2}{M1}=-\frac{M_N \omega_{cm}}{6}\, \frac{r_n^2}{2 \mu_n},
\ee
where the last expression is obtained in the zero momentum transfer limit.
Here, $r_n^2$ is the neutron charge radius (in fm$^2$), 
$\mu_n$ the neutron magnetic moment (divided by $\mu_N=e/(2 M_N)$), 
and $\omega_{cm}$ is the center of mass
energy of the photon-nucleon system at the $\Delta$ resonance. 
Fig.\ref{fig:ratio} shows the prediction~\cite{Gra00} of $C2/M1$ based on 
the elastic neutron form factor data and Eq.(\ref{c2m1}). 
Sign and magnitude of the $C2/M1$ ratio calculated in this way 
are in astonishingly good agreement with direct pion 
electroproduction data~\cite{Bar99,c2old}. This supports our finding 
that the neutron and $N \to \Delta$ 
quadrupole transition form factors are related as suggested by 
Eq.(\ref{c2m1}).

\begin{figure}[htb]
\hspace{-0.1 cm} 
\epsfxsize 12.3 true cm   
\epsfysize 5.8 true cm 
\epsfbox{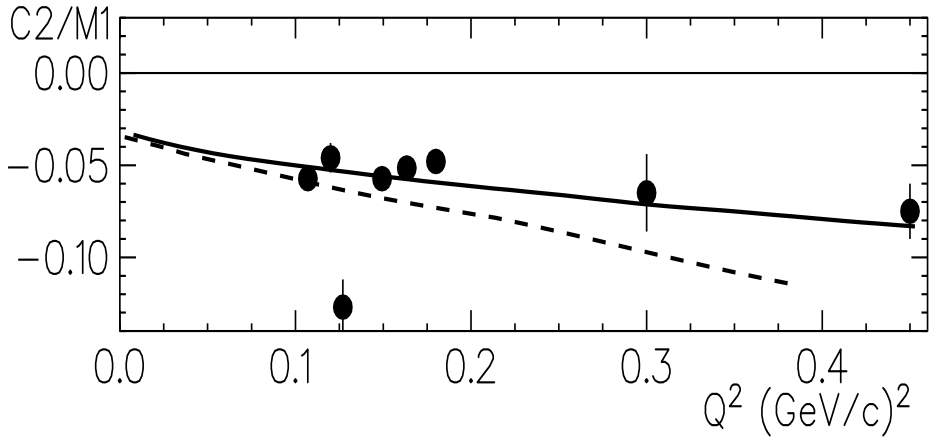}
\caption{\label{fig:ratio}
  Ratio $C2/M1$ from the fit to \Gen\ (solid curve)~\protect\cite{Gra00} 
and from
  calculations of Ref.~\protect\cite{Buc00a} (dashed curve) in comparison with
  experimental results taken from Refs.~\protect\cite{Bar99,c2old}.}
\end{figure}

\goodbreak 
\section{Summary}
\nobreak

By including two-body currents 
in the constituent quark model we have found a number of new relations 
between the elastic electromagnetic form factors of the $N$ and $\Delta$.
These relations contain the effect of the SU(6) symmetry breaking
spin-spin term in the charge operator of Eq.(\ref{eq:structure}). 
The latter lifts the degeneracy of the nucleon and $\Delta$
charge form factors and shows that their difference 
is equal to the neutron charge form factor.
Dillon and Morpurgo have recently proven that the ensuing 
Eq.(\ref{crdn}), which relates proton, neutron, and $\Delta$ charge
radii is generally valid if the (numerically small) 
three-quark operators and strange quark loops are neglected.
Furthermore, in a large $N_c$  approach we find that 
the relation $G_C^p(Q^2) - G_C^{\Delta^+}(Q^2)= 
G_C^n(Q^2) - G_C^{\Delta^0}(Q^2)$ holds 
{\it for any $N_c$ } and that it is 
broken only by small three-body $O(1/N_c^2)$ terms, which also
underlines its generality.

The $N \to \Delta$ quadrupole transition form factor $G_{C2}^{p \to \Delta}$
is found to be expressable in terms of the neutron charge form factor 
$G_C^n$. Eq.(\ref{qnff}) displays the underlying $SU(6)$ symmetry and
its breaking due to the spin-dependent two-body exchange currents.
The $SU(6)$ relation between the
$N \to \Delta$ magnetic dipole transition form factor 
and the magnetic neutron form factor is not changed 
by the two-body exchange currents of Fig.\ref{fig:feynmec}.
Even after the inclusion of spatial two-body currents the 
$N \to \Delta$ transition magnetic moment is 
given by the Beg-Lee-Pais relation.

In summary, our theory leads to a number of new relations 
between the electromagnetic form factors
of the nucleon and the $\Delta$. In particular,  
the $C2/M1$ ratio in the electromagnetic $N \to \Delta$ transition 
is given by the ratio of the elastic neutron charge and magnetic form factors.
The good agreement between our prediction for the $C2/M1$ ratio and the 
pion electroproduction data supports our analysis. 
Our work makes it clear 
that the deformation of the nucleon and the 
neutron charge radius are 
related phenomena. They are different manifestations
of the $q\bar q$ degrees of freedom in the nucleon, which are 
in leading order expressable by two-body exchange currents.

\noindent 
{\bf Acknowledgement:} This work is supported by the DFG BU813/2-1 
and Jefferson Lab.}

\eject


\vspace*{-2pt}

\begin{thebibliography}{99}
\bibitem{Buc91}  A. Buchmann, E. Hern\'andez, and K. Yazaki, 
\Journal{\PLB}{269}{35}{1991};
\Journal{\NPB}{569}{661}{1994}.
\bibitem{Gra00} P. Grabmayr and A. J. Buchmann,
\Journal{\PRL}{86}{2237}{2001}. 
\bibitem{Buc97} A. J. Buchmann, E. Hern\'andez, and A. Faessler,
\Journal{\PRC}{55}{448}{1997}.
\bibitem{Col85} S. Coleman, Aspects of Symmetry, Cambridge University Press,
1985.
\bibitem{Buc97b} A. J. Buchmann, Z. Naturforsch. {\bf 52a}, 877 (1997).
\bibitem{Buc00a} A. J. Buchmann, \Journal{\NPA}{670}{174}{2000}.
\bibitem{Mor99} G. Dillon and G. Morpurgo, 
\Journal{\PLB}{448}{107}{1999}. 
\bibitem{Buc00b} A. J. Buchmann and R. F. Lebed,  
\Journal{\PRD}{62}{096005}{2000}.
\bibitem{Buc00c} A. J. Buchmann and E. M. Henley, 
\Journal{\PRC}{63}{001}{2001}.
\bibitem{Her99} 
C. Herberg {\it et} al., Eur. Phys. J.{\bf A5} (1999) 131; J. Becker 
{\it et} al., ibid.; 
M. Ostrick {\it et} al., \Journal{\PRL}{83}{276}{1999}; 
I. Passchier {\it et}  al., \Journal{\PRL}{82}{4988}{1999}; 
for refs. to the other data see Ref.\cite{Buc97b}.
\bibitem{PDG94} L. Montanet {\it et} al., 
\Journal{\PRD}{50}{1173}{1994}.  
\bibitem{Bec97} R. Beck {\it et} al., \Journal{\PRL}{78}{606}{1997};
 G. Blanpied {\it et} al., \Journal{\PRL}{79}{4337}{1997}.
\bibitem{Beg64} M. A. B. Beg, B. W. Lee and A. Pais,
\Journal{\PRL}{13}{514}{1964}.  
\bibitem{Mor89} G. Morpurgo, \Journal{\PRD}{40}{2997}{1989}.  
\bibitem{Buc99} A. J. Buchmann and E. M. Henley,
\Journal{\PLB}{484}{255}{2000}. 
\bibitem{kopecky}  S. Kopecky \etal, \Journal{\PRL}{74}{2427}{1995} 
\bibitem{Bar99}  P. Bartsch {\it et} al.,
Baryons' 98, Bonn, Germany, World Scientific, Singapore, 1999,
 eds. D. W. Menze and B. Metsch, pg. 757; M. O. Distler, ibid., 
pg. 753; R. Gothe, ibid. pg. 394, and these proceedings.
\bibitem{c2old} R. Siddle \etal, \Journal{\NPB}{35}{93}{1971};  
 J.C. Alder \etal, \Journal{\NPB}{46}{573}{1972}.  
\end{thebibliography}
\end{document}